\documentclass[%
 reprint,
superscriptaddress,
 amsmath,amssymb,
 aps,
]{revtex4-2}

\usepackage{graphicx}
\usepackage{dcolumn}
\usepackage{amsmath}
\usepackage{bm}
\usepackage{xcolor}
 \usepackage{subcaption}

\begin{document}

\preprint{APS/123-QED}

\title{Post-selective attack with multi-mode projection onto Fock subspace}

\author{Andrei Gaidash}
\affiliation{Laboratory of Quantum Processes and Measurements, ITMO University, 3b Kadetskaya Line, 199034 Saint Petersburg, Russia and SMARTS-Quanttelecom LLC, 199178, 59-1-B 6th line of Vasilievsky island, Saint Petersburg, Russia}

\author{George Miroshnichenko}
\affiliation{%
 Waveguide Photonics Research Center, ITMO University, 49 Kronverksky Prospekt, 197101 Saint Petersburg, Russia and Institute “High School of Engineering,” ITMO University, 49 Kronverksky Prospekt, 197101 Saint Petersburg, Russia
}%

\author{Anton Kozubov}
 \affiliation{Laboratory of Quantum Processes and Measurements, ITMO University, 3b Kadetskaya Line, 199034 Saint Petersburg, Russia and SMARTS-Quanttelecom LLC, 199178, 59-1-B 6th line of Vasilievsky island, Saint Petersburg, Russia}

\date{\today}

\begin{abstract}
In this work we present a comprehensive analysis of a post-selective attack on quantum key distribution protocols employing phase-encoded linearly independent coherent states (or similar alternatives). The attack relies on multimode projection onto a Fock subspace and enables probabilistic extraction of information by an eavesdropper.  We derive analytical expressions for the information accessible to the adversary and show that it depends only on three protocol parameters: the mean photon number of the signal states, the phase separation in the information basis, and the expected optical loss of the quantum channel. Several optical realizations of phase-encoded quantum key distribution protocols are analyzed to illustrate the applicability of the results. Possible countermeasures against the proposed attack are also discussed.
\end{abstract}

\maketitle


\section{Introduction}\label{sec-intro}

Quantum key distribution (QKD) protocols derive their security from the fundamental laws of quantum physics \cite{pirandola2020advances}. Over the past decades, substantial progress has been made in the theoretical analysis of QKD security. While initial security analyses were based on the assumption of ideal components and operations, subsequent research has shifted toward more practical models. In recent years, the focus has moved toward understanding how imperfections in real-world devices impact the secrecy and reliability of QKD systems \cite{gaidash2022subcarrier,sajeed2021approach,wang2018finite,pereira2019quantum,wang2021measurement,navarrete2021practical,makarov2024preparing,fadeev2025optical}. 

State preparation plays a central role in QKD security, as it defines the information potentially available to an eavesdropper and constrains optimal attack strategies. While the original BB84 protocol was conceived for single-photon sources, most modern QKD systems rely on attenuated laser radiation. The resulting Poissonian photon-number distribution exposes the system to photon-number-splitting (PNS) attacks \cite{huttner1995coherent,lutkenhaus2000security,lutkenhaus2002quantum,acin2004pns}. A widely adopted and highly effective countermeasure against this class of attacks is the decoy-state technique \cite{lo2005decoy,ma2005practical,wang2005beating}, including phase randomization and single-photon fraction extraction, also known as the Gottesman-Lo-Lutkenhaus-Preskill (GLLP) framework \cite{gottesman2004security}, at the processing step of a protocol. 

However, specifically for phase-coded QKD protocols that will be considered in the article, a generic PNS attack is generally considered less critical than other eavesdropping strategies, as the encoded phase is fragile with respect to photon-number measurements, i.e., the direct projection of a single-mode coherent state on Fock subspace erases information about the phase; see Section 3.2 in \cite{chistiakov2019feasibility}. 
At this point, consideration of collective attacks, e.g., the Holevo bound, for phase-coded states was assumed to be optimal (implying decoy states and single-photon fraction extraction at the processing step of a protocol are not implemented); for instance, see \cite{sajeed2021approach}. Then, a thorough analysis of unambiguous state discrimination (that can be implemented for a set of linearly independent states) attacks shows that a post-selection (by blocking undesired outcomes) can, in principle, surpass collective attacks' estimations, and the only way to observe an eavesdropper's intervention is to monitor the expected detection (and/or bit error) rate \cite{gaidash2019countermeasures,gaidash2022subcarrier}. As a result, the paradigm of quantum control attack has emerged \cite{kozubov2021quantum}, as well as effective countermeasures. Preventing, or at least limiting, the ability of an eavesdropper to block undesired outcomes results in the Holevo boundary being re-established as the maximal amount of information retrieved (accompanied by detection and/or bit error rate monitoring conditions), e.g., \cite{sajeed2021approach}.

Later, the formalism of post-selective measurements was developed \cite{kronberg2021increasing,kronberg2022success,kenbaev2022quantum,klevtsov2023assisted,avanesov2023postselective,kronberg2024structure,kronberg2025local,kronberg2025generalization,zhigalskii2025collective}. The central idea is that, for a given ensemble of initial pure quantum states, it is sufficient to establish the existence of a transformation that produces a desired set of output states with a non-zero probability, without requiring explicit construction of the transformation itself. In the paper \cite{kronberg2022success} necessary and sufficient conditions for the existence of such a transformation were derived as $G_{in}-p_sG_{out}\ge0$, where $G_{in}$ and $G_{out}$ are Gram matrices of input and output states and $p_s$ is the probability of the successful transformation. This formulation allows one to characterize the feasibility of a given post-selective transformation solely through algebraic constraints on the corresponding Gram matrices. A key advantage of this approach is that it circumvents the need to specify the physical realization or operational details of the transformation. Instead, it suffices to demonstrate that a transformation with the required properties exists in principle. This feature has proven particularly valuable in the security analysis of quantum key distribution protocols, where one typically assumes an adversary with unrestricted technological capabilities. However, a general optimization over all possible output states seems to be unachievable (at least for now); recent application cases, for instance \cite{kronberg2025generalization}, still consider special cases of outcome states. Despite the latter, the attack again surpasses the Holevo bound.

Ideas of post-selection continue to develop in the article with an unexpected combination of PNS-inspired techniques. In contrast to recent papers \cite{kronberg2021increasing,kronberg2022success,kenbaev2022quantum,klevtsov2023assisted,avanesov2023postselective,kronberg2024structure,kronberg2025local,kronberg2025generalization,zhigalskii2025collective}, necessary transformations of the proposed attack are fully described. The key idea of the proposed attack and the attack scenario are described in Section~\ref{sec-key-idea}. The efficiency of the attack is studied in Section~\ref{sec-efficiency} by comparison with the Holevo bound (as in \cite{kronberg2025generalization}) for different values of a protocol's characteristics. Several QKD protocols and their optical schemes are considered as application cases for the proposed attack in Section~\ref{sec-application}. Section~\ref{sec-result} concludes the article.


\section{Key idea}\label{sec-key-idea}

\subsection{Multi-mode projection onto the Fock subspace}

Consider a coherent state $|\alpha e^{i\phi_j}\rangle$, where phase $\phi_j$ can be chosen from a set of phases that are equally distributed in a phase plane. Assume that an eavesdropper concatenates a coherent state $|\alpha\rangle$ to obtain a state as follows:
\begin{gather}
    |\psi(\alpha,\phi_j)\rangle=|\alpha\rangle\otimes|\alpha e^{i\phi_j}\rangle.\label{addingmode}
\end{gather}
Note, an eavesdropper can add a phase-matched coherent state reliably since there is always a reference beam in every interferometric scheme; see further Section~\ref{sec-application}. The latter state can be projected onto the Fock subspace with the following operator:
\begin{gather}
    Q^{(n)}=\sum_{k=0}^{n}|n-k\rangle\langle n-k|\otimes|k\rangle\langle k|,\label{project}
\end{gather}
resulting in the following reduced state:
\begin{gather}
  |\xi^{(n)}(\alpha,\phi_j)\rangle=\sum_{k=0}^{n}\frac{\langle n-k|\alpha\rangle\cdot|n-k\rangle\otimes \langle k|\alpha e^{i\phi_j}\rangle\cdot|k\rangle}{\sqrt{P(2|\alpha|^2,n)}}=\notag\\
  =\sum_{k=0}^{n}\frac{e^{i(\phi_jk+\theta n)}}{2^{n/2}}\sqrt{\binom{n}{k}}|n-k\rangle\otimes|k\rangle,\label{reduced}\\
  P(2|\alpha|^2,n)=\text{Tr}\big( Q^{(n)}|\psi(\alpha,\phi_j)\rangle\langle\psi(\alpha,\phi_j)|\big),\label{poissprob}\\
  P(x,n)=\frac{e^{-x}x^n}{n!},
\end{gather}
where $\theta$ is a phase of a complex-valued coherent amplitude $\alpha=|\alpha|e^{i\theta}$, and $P(x,n)$ is the Poisson distribution. The scalar product of reduced states is given by
\begin{gather}
    \langle\xi^{(n)}(\alpha,\phi_i)|\xi^{(n)}(\alpha,\phi_j)\rangle=\bigg(\frac{1+e^{-i(\phi_i-\phi_j)}}{2}\bigg)^n,\label{reduceoverlap}
\end{gather}
the absolute value of their scalar product as follows:
\begin{gather}
    |\langle\xi^{(n)}(\alpha,\phi_i)|\xi^{(n)}(\alpha,\phi_j)\rangle|=\bigg(\cos\Big(\frac{\phi_i-\phi_j}{2}\Big)\bigg)^n.\label{absscalar}
\end{gather}
It is well known that with the help of the latter quantity, the maximal amount of information an eavesdropper can extract is determined, i.e., the Holevo bound. In particular, the scalar product should be considered, implying two states that form an informational basis, i.e., with some phases $\phi_{(0)}$ and $\phi_{(1)}$, that correspond to bit values of $0$ and $1$ respectively in that basis. A note should be made that it is so only if an eavesdropper performs collective measurement after legitimate users reveal information about base choices.
In this case the Holevo bound is given by
\begin{gather}
    \chi^{(n)}(\Delta)=h\Big(\frac{1-|\langle\xi^{(n)}(\alpha,\phi_{(0)})|\xi^{(n)}(\alpha,\phi_{(1)})\rangle|}{2}\Big)=\notag\\
    =h\Big(\frac{1-\big(\cos(\Delta)\big)^n}{2}\Big),
\end{gather}
where
\begin{gather}
    h(x)=-x \log_2(x)-(1-x)\log_2(1-x)
\end{gather}
is the binary entropy function and $\Delta=\frac{\phi_{(0)}-\phi_{(1)}}{2}$. The Holevo bound $\chi^{(n)}(\Delta)$ determines the total information retrieved from reduced states for a given photon-number outcome of the projection~\eqref{project}, that occurs with the probability~\eqref{poissprob}. This quantity is essential for the attack scenario described further.

\subsection{Attack scenario description}\label{sec-attackdescript}

\begin{figure*}[ht]
    \centering
    \includegraphics[scale = 0.65]{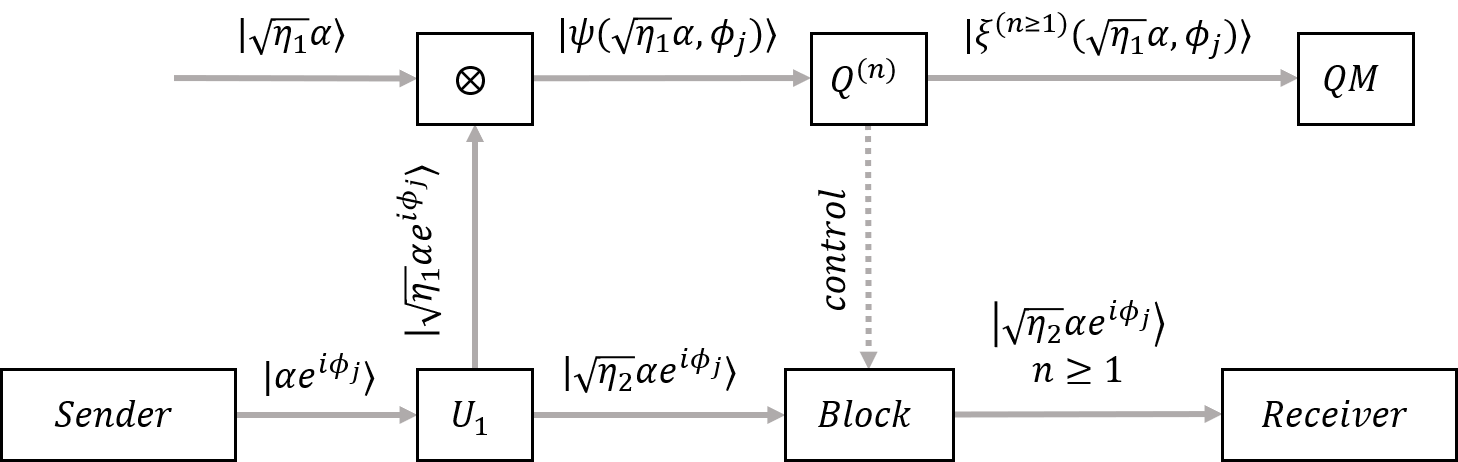}
    \caption{The principal scheme of the proposed attack, as it is described in Section~\ref{sec-attackdescript}. A sender transmits a coherent state $|\alpha e^{i\phi_j}\rangle$, that is split by unitary transformation (basically a beamsplitter) into $|\sqrt{\eta_1}\alpha e^{i\phi_j}\rangle$ and $|\sqrt{\eta_2}\alpha e^{i\phi_j}\rangle$. The former one is augmented by $|\sqrt{\eta_1}\alpha \rangle$ (made from a reference beam) to form $|\psi(\sqrt{\eta_1}\alpha,\phi_j)\rangle$ as in Eq.~\eqref{addingmode}; the latter one is transferred further to a receiver. Then the projection $Q^{(n)}$ as in Eq.~\eqref{project} is applied to $|\psi(\sqrt{\eta_1}\alpha,\phi_j)\rangle$ to form the reduced state $|\xi^{(n)}(\sqrt{\eta_1}\alpha,\phi_j)\rangle$ as in Eq.~\eqref{reduced}. If the outcome of the projection yields $n=0$, then the transferred to a receiver state is blocked (triggered by a control, denoted by a dashed arrow); otherwise ($n\ge1$), it travels further to a receiver, and the reduced state is stored in a quantum memory (QM)}
    \label{fig-scheme}
\end{figure*}

The proposed attack scenario using multi-mode projection onto the Fock subspace is as follows; also see the attack principal scheme in Fig.~\ref{fig-scheme}:

\begin{enumerate}
    \item An eavesdropper can use a beamsplitter (or a unitary operation $U_1$ in general) to split the signal in a specific way given by:
\begin{gather}
    |\alpha e^{i\phi_j}\rangle\xrightarrow{U_1}|\sqrt{\eta_1}\alpha e^{i\phi_j}\rangle\otimes|\sqrt{\eta_2}\alpha e^{i\phi_j}\rangle,\\
    |\eta_1|+|\eta_2|=1;
\end{gather}
\item The state $|\sqrt{\eta_2}\alpha e^{i\phi_j}\rangle$ is sent to a receiver while the state $|\sqrt{\eta_1}\alpha e^{i\phi_j}\rangle$ is in possession of an eavesdropper;
\item An eavesdropper proceeds with a transformation (if necessary; some of the considered further examples do not require it; for details, see Section~\ref{sec-application}) to obtain a state of the form~\eqref{addingmode}:

\begin{gather}
    |\sqrt{\eta_1}\alpha e^{i\phi_j}\rangle\rightarrow|\psi(\sqrt{\eta_1}\alpha,\phi_j)\rangle\label{addingmodeeta1};
\end{gather}
\item Then an eavesdropper applies projection~\eqref{project} to the state as follows: 
\begin{gather}
    |\psi(\sqrt{\eta_1}\alpha,\phi_j)\rangle\xrightarrow{Q^{(n)}}|\xi^{(n)}(\sqrt{\eta_1}\alpha,\phi_j)\rangle;
\end{gather}
\item Reduced states in case of $n=0$ are indistinguishable; they are basically vacuum states: $|\xi^{(0)}(\sqrt{\eta_1}\alpha,\phi_j)\rangle=|0\rangle\otimes|0\rangle$, and they do not provide any information in regard to the sent state's phase $\phi_j$. So, at this point, an eavesdropper may apply post-selection (for instance, implemented as in \cite{kozubov2021quantum}):
\begin{itemize}
    \item If $n=0$, then an eavesdropper blocks the state $|\sqrt{\eta_2}\alpha e^{i\phi_j}\rangle$ sent to a receiver, and the reduced state $|\xi^{(0)}(\sqrt{\eta_1}\alpha,\phi_j)\rangle$ is neglected;
    \item If $n\ge1$, then the state $|\sqrt{\eta_2}\alpha e^{i\phi_j}\rangle$ proceeds to a receiver's side, and the reduced state $|\xi^{(n\ge1)}(\sqrt{\eta_1}\alpha,\phi_j)\rangle$ is stored in a quantum memory.
\end{itemize}
\item After the reconciliation step, when informational bases are revealed, each stored in a quantum memory state results in information, quantified by $\chi^{(n)}(\Delta)$, obtained by an eavesdropper. Then, the total information about a key retrieved from reduced states, excluding vacuum states, is given by
\begin{gather}
    I=\sum_{n=1}^{\infty}\frac{P(2|\eta_1||\alpha|^2,n)\cdot\chi^{(n)}(\Delta)}{1-P(2|\eta_1||\alpha|^2,0)},
\end{gather}
where we utilize the following property of von Neumann entropy $H$:
\begin{gather}
    H\Big(\bigotimes_{j}\rho_j\Big)=\sum_jH(\rho_j),
\end{gather}
and normalize the quantity with respect to blocked events ($n=0$).
\end{enumerate}

\section{Parameters adjustment and efficiency estimations}\label{sec-efficiency}
\subsection{Splitting coefficients}
Note that splitting coefficients $\eta_1$ and $\eta_2$ (more precisely, their absolute values since a phase does not contribute) should be adjusted in a specific way. In particular, the detection rate (for the sake of simplicity, we consider the Mandel approximation) at a receiver's side with post-selection should be unchanged compared to the case without the attack; the latter results in the following equation:
\begin{gather}
    \big(1-P(2|\eta_1||\alpha|^2,0)\big)|\eta_2|=\eta_L,\label{cond1}
\end{gather}
where real-valued $\eta_L$ is the expected attenuation of a signal in a quantum channel with length $L$. Considering $P(2|\eta_1||\alpha|^2,0)=e^{-2|\eta_1||\alpha|^2}\approx1-2|\eta_1||\alpha|^2$ and $|\eta_2|=1-|\eta_1|$, the latter equation results in:
\begin{gather}
    2|\eta_1||\alpha|^2(1-|\eta_1|)=\eta_L,
\end{gather}
then
\begin{gather}
    |\eta_1|=\frac{1}{2}\Big(1\pm\sqrt{1-\frac{2\eta_L}{|\alpha|^2}}\Big),\quad |\eta_2|=\frac{1}{2}\Big(1\mp\sqrt{1-\frac{2\eta_L}{|\alpha|^2}}\Big).
\end{gather}

The latter equations for $|\eta_1|$ and $|\eta_2|$ show that the attack can be performed only if $\eta_L\le\frac{|\alpha|^2}{2}$ (the root term should be positive), i.e., in a low transmission or, equivalently, high loss region; otherwise, an eavesdropper cannot maintain the detection rate. In order to perform an attack in the region $\eta_L\ge\frac{|\alpha|^2}{2}$, the easiest way is to apply the attack only to a fraction of optical signals; the other part of optical signals is to be transferred to a receiver without any losses. Then condition~\eqref{cond1} can be replaced by
\begin{gather}
    z\big(1-P(2|\eta_1||\alpha|^2,0)\big)|\eta_2|+(1-z)=\eta_L,
\end{gather}
where $z\le 1$ denotes the fraction of attacked signals. Applying the same approximations results in:
\begin{gather}
    |\eta_1|=\frac{1}{2}\Big(1\pm\sqrt{1-\frac{2(\eta_L-(1-z))}{z|\alpha|^2}}\Big),\label{eta1eq}\\
    |\eta_2|=\frac{1}{2}\Big(1\mp\sqrt{1-\frac{2(\eta_L-(1-z))}{z|\alpha|^2}}\Big).
\end{gather}
The maximal value of $z$ in the region $\eta_L\ge\frac{|\alpha|^2}{2}$ is as follows:
\begin{gather}
    z=\frac{1-\eta_L}{1-\frac{|\alpha|^2}{2}}.
\end{gather}
Then, for any value of $\eta_L$, fraction $z$ can be generalized as
\begin{gather}
    z=\min\Big(1,\frac{1-\eta_L}{1-\frac{|\alpha|^2}{2}}\Big).\label{zeq}
\end{gather}
We summarize as follows:
\begin{itemize}
    \item If $\eta_L\ge\frac{|\alpha|^2}{2}$, then $z=\frac{1-\eta_L}{1-\frac{|\alpha|^2}{2}}$, and $2 |\eta_1| =1$;
    \item If $\eta_L\le\frac{|\alpha|^2}{2}$, then $z=1$, and $|\eta_1|$ should be defined as in Eq.~\eqref{eta1eq}.
\end{itemize}

\subsection{Attack efficiency}
At this point, total information obtained by an eavesdropper can be estimated by
\begin{gather}
    I=z\sum_{n=1}^{\infty}\frac{P(2|\eta_1||\alpha|^2,n)\cdot\chi^{(n)}(\Delta)}{1-P(2|\eta_1||\alpha|^2,0)},\label{info}
\end{gather}
where $z$ is defined in Eq.~\eqref{zeq} and $|\eta_1|$ is defined in Eq.~\eqref{eta1eq}. Analysis of the equation shows that $|\eta_1|$ with a plus sign results in slightly more information to be extracted. Hence, hereinafter we assume $|\eta_1|$ in Eq.~\eqref{eta1eq} to be with a plus sign.

\begin{figure}
\includegraphics[width=1\linewidth]{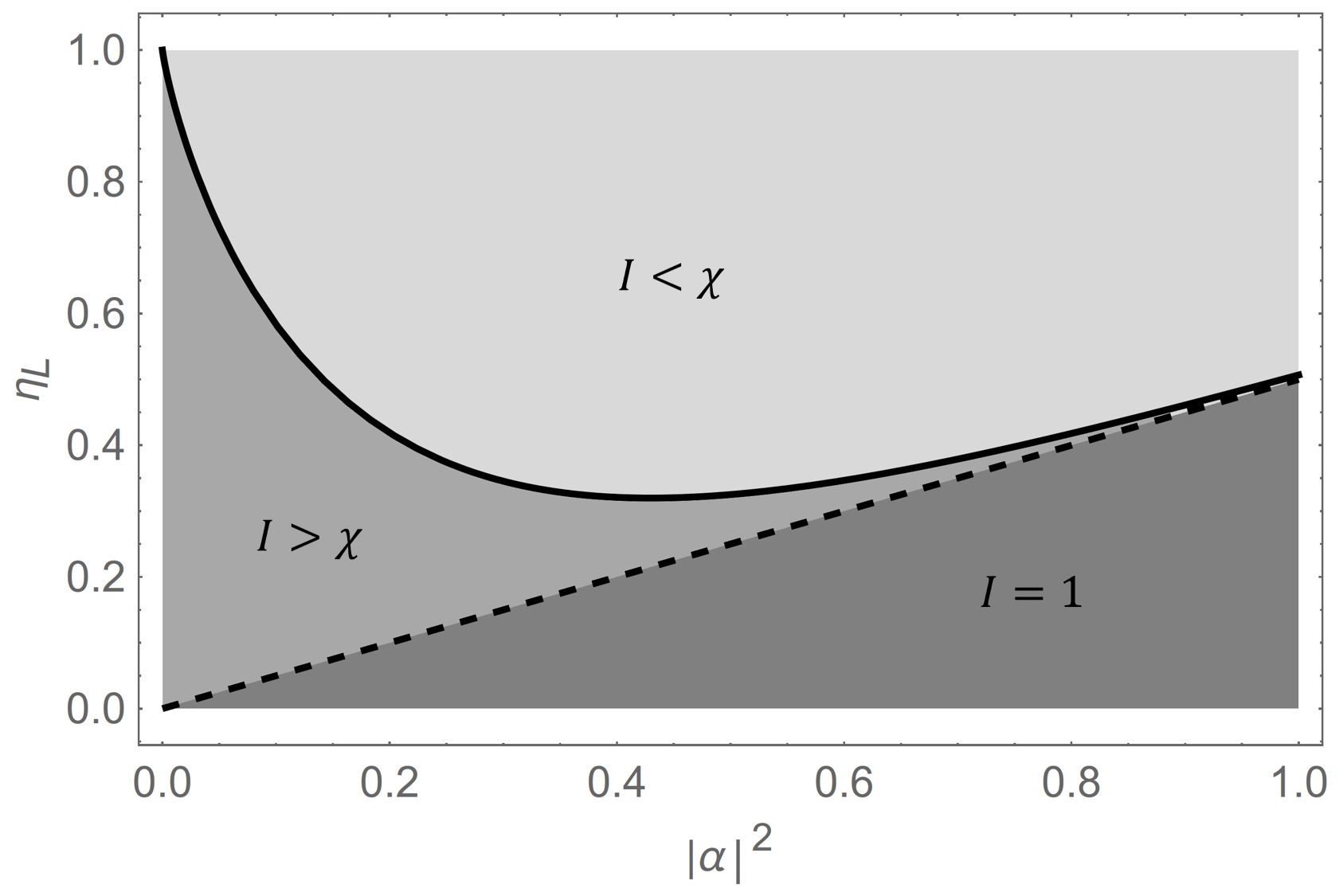}
\caption{Comparison of the information obtained by performing the proposed attack, denoted by $I$ and expressed in Eq.~\eqref{info} with the one obtained by performing a general collective attack, i.e., the Holevo bound, denoted by $\chi$ and expressed in Eq.~\eqref{holevo}, assuming $\Delta=\pi/2$. Three regions are shown: 1) the top region (light gray) defined by $\eta_L\ge1-\Big(1-\frac{|\alpha|^2}{2}\Big)\chi$, where $I<\chi$, 2) the middle region (gray) defined by $\frac{|\alpha|^2}{2}\le\eta_L\le1-\Big(1-\frac{|\alpha|^2}{2}\Big)\chi$, where $I>\chi$, and 3) the bottom region (dark gray) defined by $\eta_L\le\frac{|\alpha|^2}{2}$, where $\chi<I=1$}
\label{fig:holevocomp} 
\end{figure}

A common feature of QKD protocols with linearly independent states (also see Section~\ref{sec-application}) is that the maximal amount of eavesdropper's information is supposed to be limited by the Holevo bound, see App.~\ref{app1}:
\begin{gather}
    \chi=h\Big(\frac{1-e^{-|\alpha|^2(1-\cos(2\Delta))}}{2}\Big).\label{holevo}
\end{gather}
So, the attack efficiency can be estimated by comparing the amount of information retrieved by the proposed attack \eqref{info} with the commonly used Holevo bound \eqref{holevo} in the same way as it was performed for a post-selective attack in \cite{kronberg2025generalization}. The other thing to note is that $\Delta=\pi/2$ for considered protocols with linearly independent states. Therefore, results of efficiency estimations are shown in Fig.~\ref{fig:holevocomp}, where three regions are as follows:
\begin{enumerate}
    \item The top region is defined by $\eta_L\ge1-\Big(1-\frac{|\alpha|^2}{2}\Big)\chi$,
    where the information extracted by an eavesdropper performing proposed attack is less than the Holevo bound calculated for the initial states in informational basis, i.e., the information obtained by an eavesdropper performing general collective attack ($I<\chi$);
    \item The middle region is defined by $\frac{|\alpha|^2}{2}\le\eta_L\le1-\Big(1-\frac{|\alpha|^2}{2}\Big)\chi$, where the proposed attack allows obtaining more information than the Holevo bound but not the entire key;
    \item The bottom region is defined by $\eta_L\le\frac{|\alpha|^2}{2}$; this is the most efficient one, where an eavesdropper possesses all information about a key ($I=1$).
\end{enumerate}
Results clearly indicate that the proposed attack is efficient for a wide range of parameters, especially in the region $|\alpha|^2\le0.1$, where most of the actual values of $|\alpha|^2$ are for practical implementations.

\section{Application cases}\label{sec-application}

In this section, several application cases are considered. We inspect utilized quantum states to be similar to the one in Eq.~\eqref{addingmode} and apply minor tweaks to the multi-mode projection operator if necessary. These features guarantee the applicability of the attack scenario and estimations described in previous sections.

\subsection{Mach-Zehnder interferometer-based}

In this section, we will consider a QKD protocol based on the optical scheme with two Mach-Zehnder interferometers (MZI) \cite{han2005stability, song2020phase}; the first one is for the state preparation, and the second one is for the interference. As we have stated earlier, we are mostly interested in the similarity of utilized states to the one in Eq.~\eqref{addingmode}. In the MZI, the coherent state is split into two paths: the short one and the long one (with a time delay). In one of them, a phase modulator is placed. After the modulation, two paths are combined. The time-delay is long enough to consider two coherent states as separate states that are identical to the one in Eq.~\eqref{addingmode}:
\begin{gather}
    |\psi_{MZI}(\alpha,\phi_j)\rangle=|\alpha\rangle\otimes|\alpha e^{i\phi_j}\rangle.\label{mzistate}
\end{gather}
At this point, the proposed attack scenario and all the estimations can be directly implemented in regard to a QKD protocol based on the MZI optical scheme.

Note that no additional states should be considered (decoy states) with different values of $|\alpha|^2$; we will catch on to that in Section~\ref{sec-result}. However, the phase randomization is not an issue for the proposed attack: there is always a reference beam that matches a global phase with a signal beam.

\subsection{Phase-time coding}

An alternative to the basic MZI scheme can be a protocol with phase-time coding that utilizes states of the following form in two ($L$ or $R$) time-slot bases \cite{molotkov2008cryptographic,molotkov2010phase}:
\begin{gather}
    |\psi_{PT}(\alpha,\phi_j)\rangle_L=|\alpha\rangle_1\otimes|\alpha e^{i\phi_{j}}\rangle_2\otimes|0\rangle_3,\label{ptstate1}\\
    |\psi_{PT}(\alpha,\phi_j)\rangle_R=|0\rangle_1\otimes|\alpha e^{i\phi_{j}}\rangle_2\otimes|\alpha\rangle_3.\label{ptstate2}
\end{gather}
In this case, an eavesdropper applies the same strategy, with the only exception of slightly modified operators for projection on a Fock subspace suitable for the latter states:
\begin{gather}
    Q_{PT}^{(n)}=\sum_{k=0}^{n}\sum_{m=0}^{k}|n-k-m\rangle\langle n-k-m|\otimes\\ \otimes|k-m\rangle\langle k -m|\otimes|m\rangle\langle m|.\nonumber
\end{gather}
Despite a small difference in the projection operators, reduced states have the same overlap as in Eq.~\eqref{reduceoverlap}:
\begin{gather}
    |\psi_{PT}(\alpha,\phi_j)\rangle_L\xrightarrow{Q_{PT}^{(n)}}|\xi^{(n)}_{PT}(\alpha,\phi_j)\rangle_L,\\
    |\psi_{PT}(\alpha,\phi_j)\rangle_R\xrightarrow{Q_{PT}^{(n)}}|\xi^{(n)}_{PT}(\alpha,\phi_j)\rangle_R,\\
    _L\langle\xi^{(n)}_{PT}(\alpha,\phi_i)|\xi^{(n)}_{PT}(\alpha,\phi_j)\rangle_L\ =\notag\\ =\ _R\langle\xi^{(n)}_{PT}(\alpha,\phi_i)|\xi^{(n)}_{PT}(\alpha,\phi_j)\rangle_R=\notag\\
    =\bigg(\frac{1+e^{-i(\phi_i-\phi_j)}}{2}\bigg)^n,
\end{gather}
and again, the proposed attack scenario can be applied to a QKD protocol based on the phase-time coding scheme in the same way.

\subsection{Phase-matching}

The recently developed phase-matching QKD protocol \cite{ma2018phase} provides an alternative to twin-field quantum key
distribution \cite{lucamarini2018overcoming}, where the secure key rate scales
with the square root of the transmission probability. However, we have found that the proposed attack strategy can be applied to the basic concept of the protocol, implying the absence of optical pulses' power change as in the decoy-state method. However, we will touch on that later in the discussion of Section~\ref{sec-result}.

Two senders prepare the following state:
\begin{gather}
    \big||\alpha| e^{i(\theta_a+\phi_a+\pi p_a)}\big\rangle\otimes \big||\alpha| e^{i(\theta_b+\phi_b+\pi p_b)}\big\rangle,\label{pmstate}
\end{gather}
where $\theta_j$ is a constant unknown global phase for each sender to be adjusted to observe interference with high visibility, $\phi_j$ is a random phase-shift that is announced during reconciliation, $p_j$ is a key bit (either $0$ or $1$), and indices $a$ and $b$ denote each sender respectively. For the sake of simplicity, hereinafter we will consider an eavesdropper's actions only for one sender, implying the same for another sender. An eavesdropper prepares an additional mode with a phase-locked coherent state $\alpha_e=|\alpha|e^{i\theta_e}$, where $\theta_e$ may not even be correlated with the global phase $\theta_j$, and concatenates it with the sender's signal:
\begin{gather}
    |\psi(\alpha,\theta_e,\theta_j,\phi_j,p_j)\rangle=\big||\alpha| e^{i\theta_e}\big\rangle\otimes\big||\alpha| e^{i(\theta_j+\phi_j+\pi p_j)}\big\rangle.\label{pmstate}
\end{gather}
Then, the reduced state has the following form:
\begin{gather}
  |\xi^{(n)}_{PM}(\alpha,\theta_e,\theta_j,\phi_j,p_j)\rangle
  =\notag\\
  =\sum_{k=0}^{n}\frac{e^{i\theta_e n}e^{i(\theta_j+\phi_j+\pi p_j-\theta_e)k}}{2^{n/2}}\sqrt{\binom{n}{k}}|n-k\rangle\otimes|k\rangle,
\end{gather}
and if $\theta_j=\theta_a=\theta_b$ is fixed, $\theta_e$ is fixed, and $\phi_j=\phi_a=\phi_b$ are announced, then the reduced states with different values of $p_j$ are orthogonal (except $n=0$):
\begin{gather}
    \langle\xi^{(n)}_{PM}(\alpha,\theta_e,\theta_j,\phi_j,0)|\xi^{(n)}_{PM}(\alpha,\theta_e,\theta_j,\phi_j,1)\rangle=\delta_{n,0}.
\end{gather}
Orthogonality of reduced states implies that the proposed attack strategy can be implemented with only one thing to note: in contrast to previously considered protocols, failure probability resulting in blockade of the optical signals is less and equals $P(2|\eta_1||\alpha|^2,0)^2$, since there are two senders: successful performance of the attack in one arm of the optical scheme results in obtaining information coded in the other arm due to a result of the interference.

\begin{figure*}
\centering
      
    \begin{subfigure}[b]{0.45\textwidth}
        \centering
        \includegraphics[width=\linewidth]{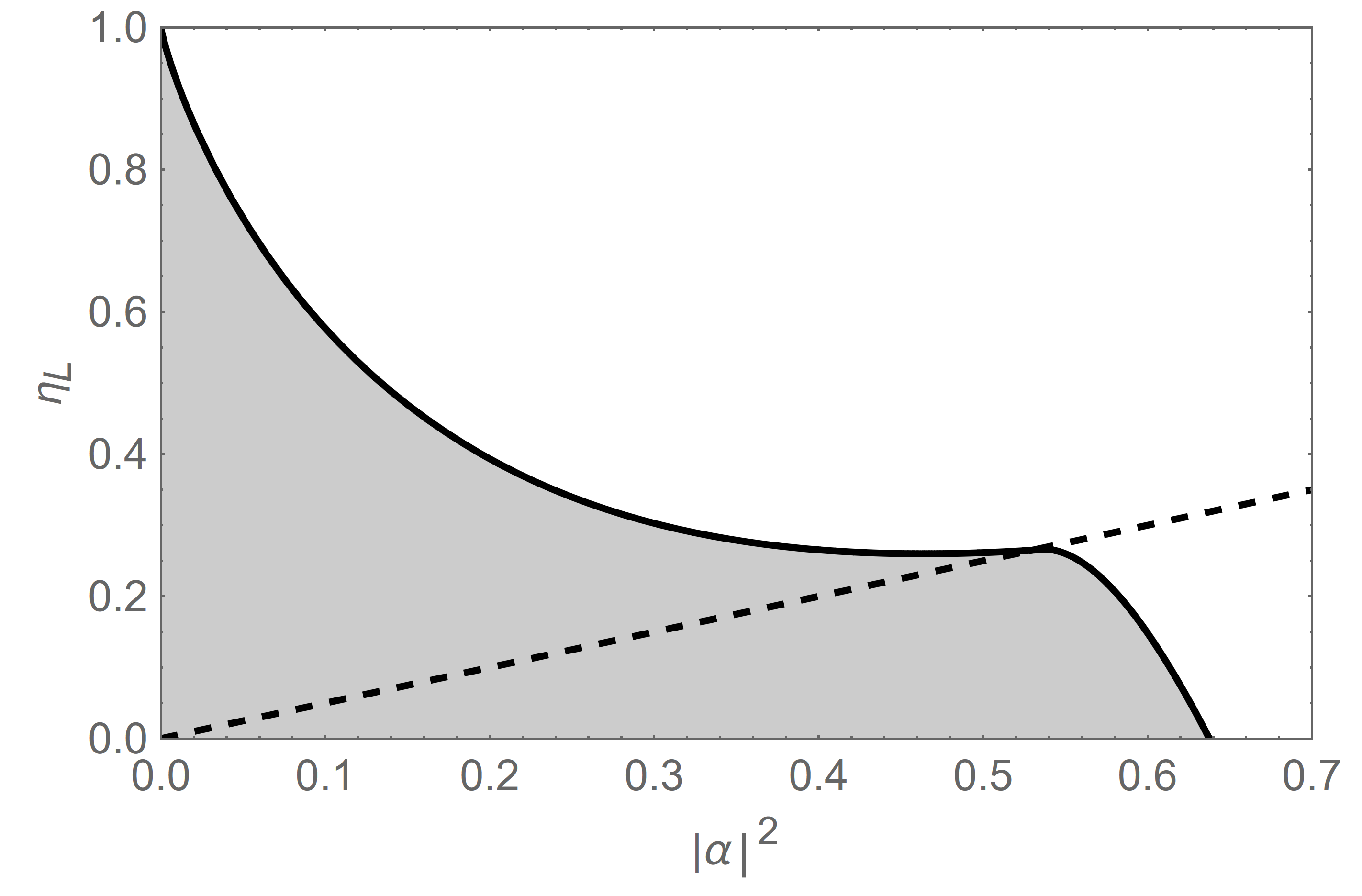}
        \caption{$\Delta=\pi/3$}
        \label{fig:fig4a}
    \end{subfigure}
  \hfill
    \begin{subfigure}[b]{0.45\textwidth}
        \centering
        \includegraphics[width=\linewidth]{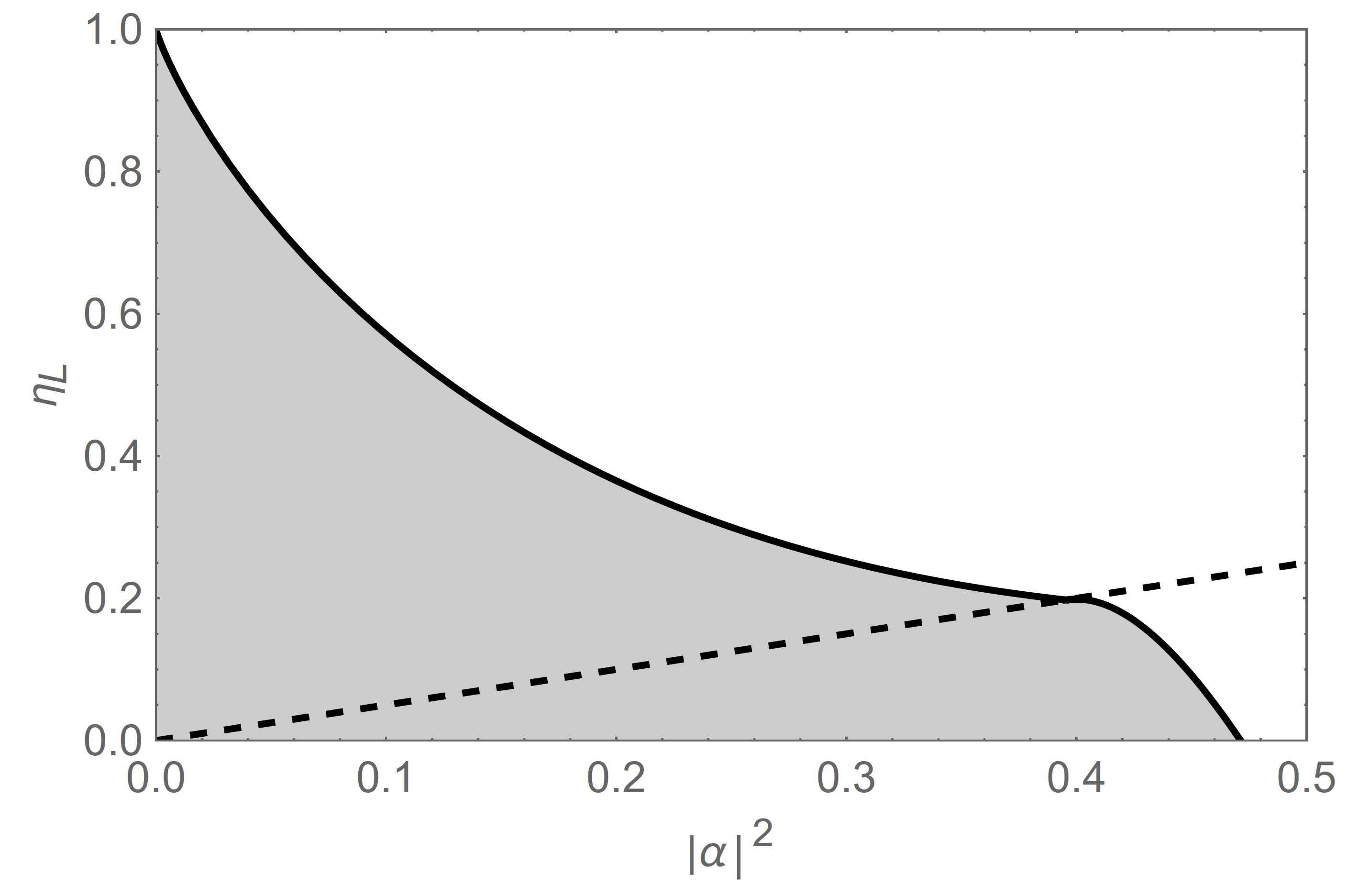}
        \caption{$\Delta=\pi/4$}
        \label{fig:fig4b}
    \end{subfigure}
    \hfill
    \begin{subfigure}[b]{0.45\textwidth}
        \centering
        \includegraphics[width=\linewidth]{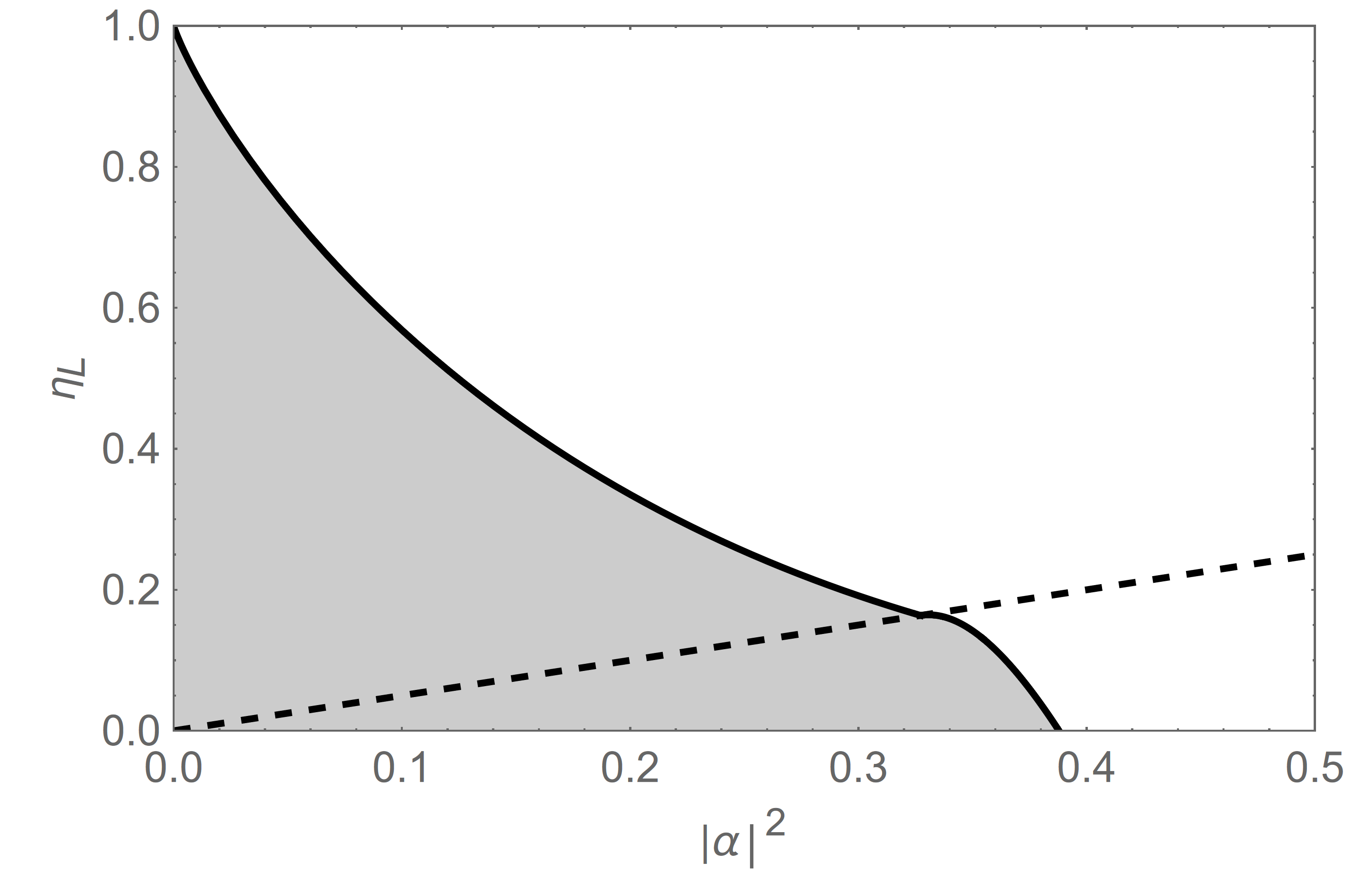}
        \caption{$\Delta=\pi/6$}
        \label{fig:fig4c}
    \end{subfigure}
    \hfill
    \begin{subfigure}[b]{0.45\textwidth}
        \centering
        \includegraphics[width=\linewidth]{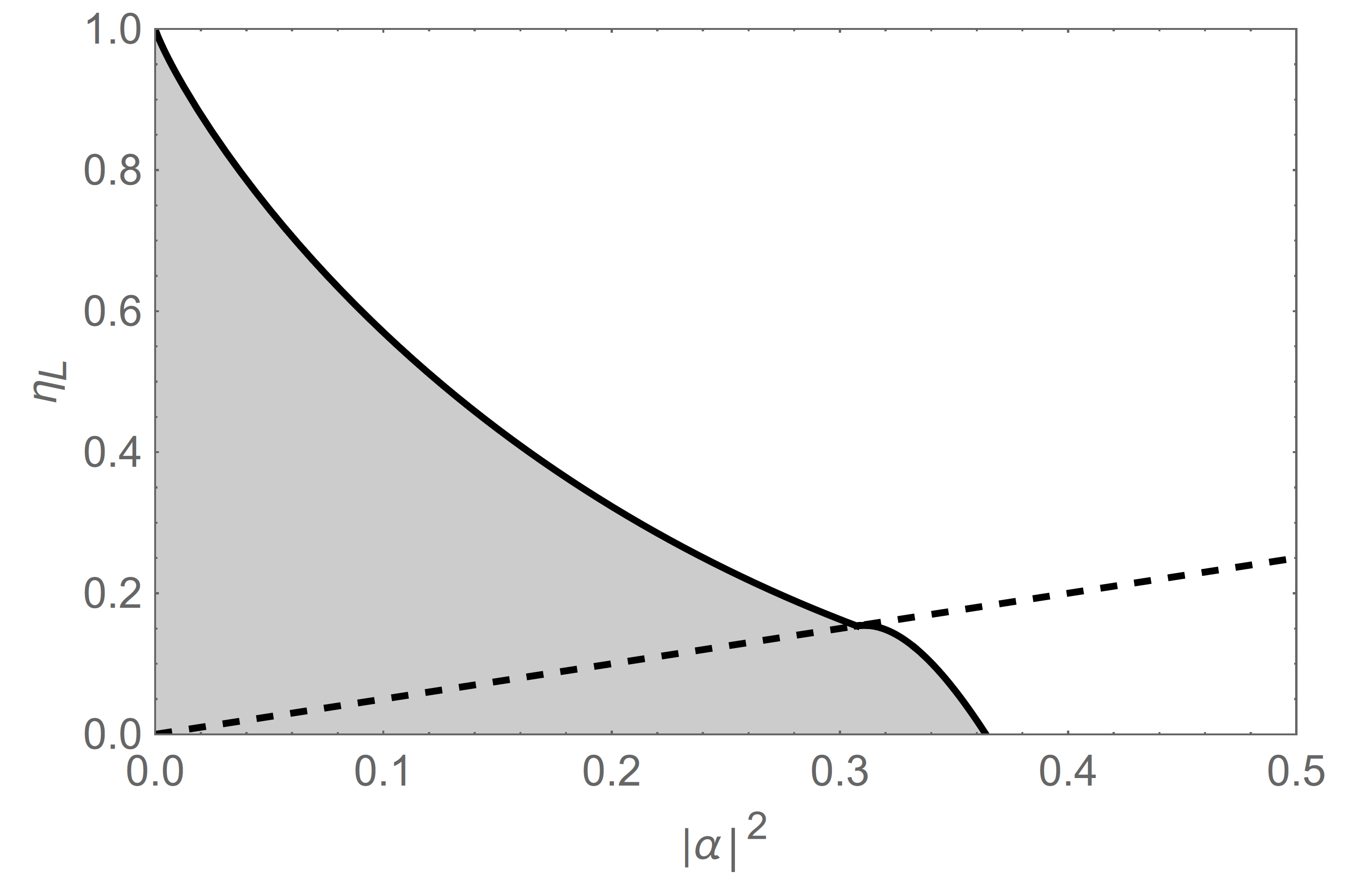}
        \caption{$\Delta=\pi/8$}
        \label{fig:fig4d}
    \end{subfigure}
    \hfill
    \begin{subfigure}[b]{0.45\textwidth}
        \centering
        \includegraphics[width=\linewidth]{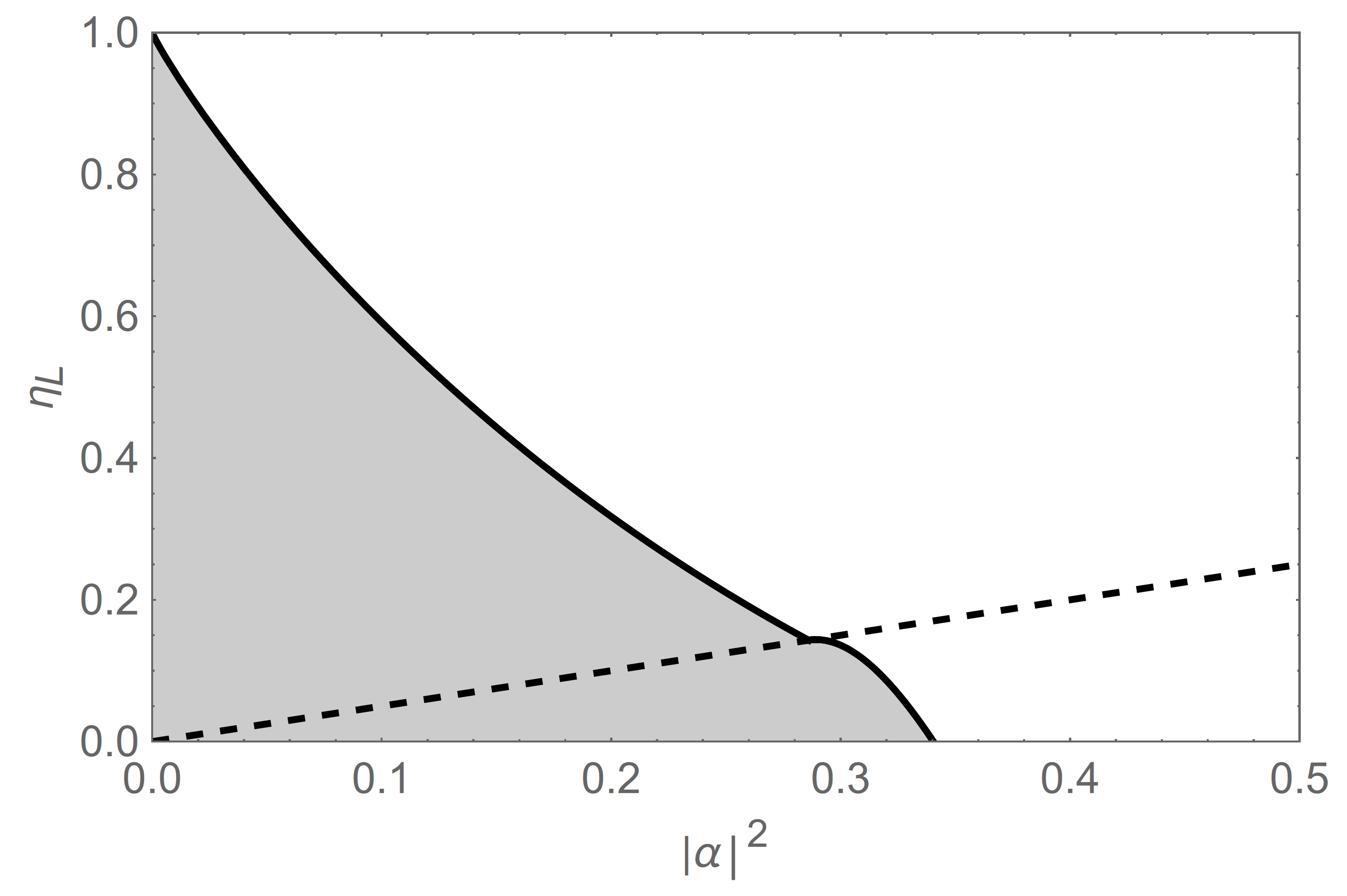}
        \caption{$\Delta\approx0.0237\cdot\pi$}
        \label{fig:fig4e}
    \end{subfigure}
    \hfill
    \begin{subfigure}[b]{0.45\textwidth}
        \centering
        \includegraphics[width=\linewidth]{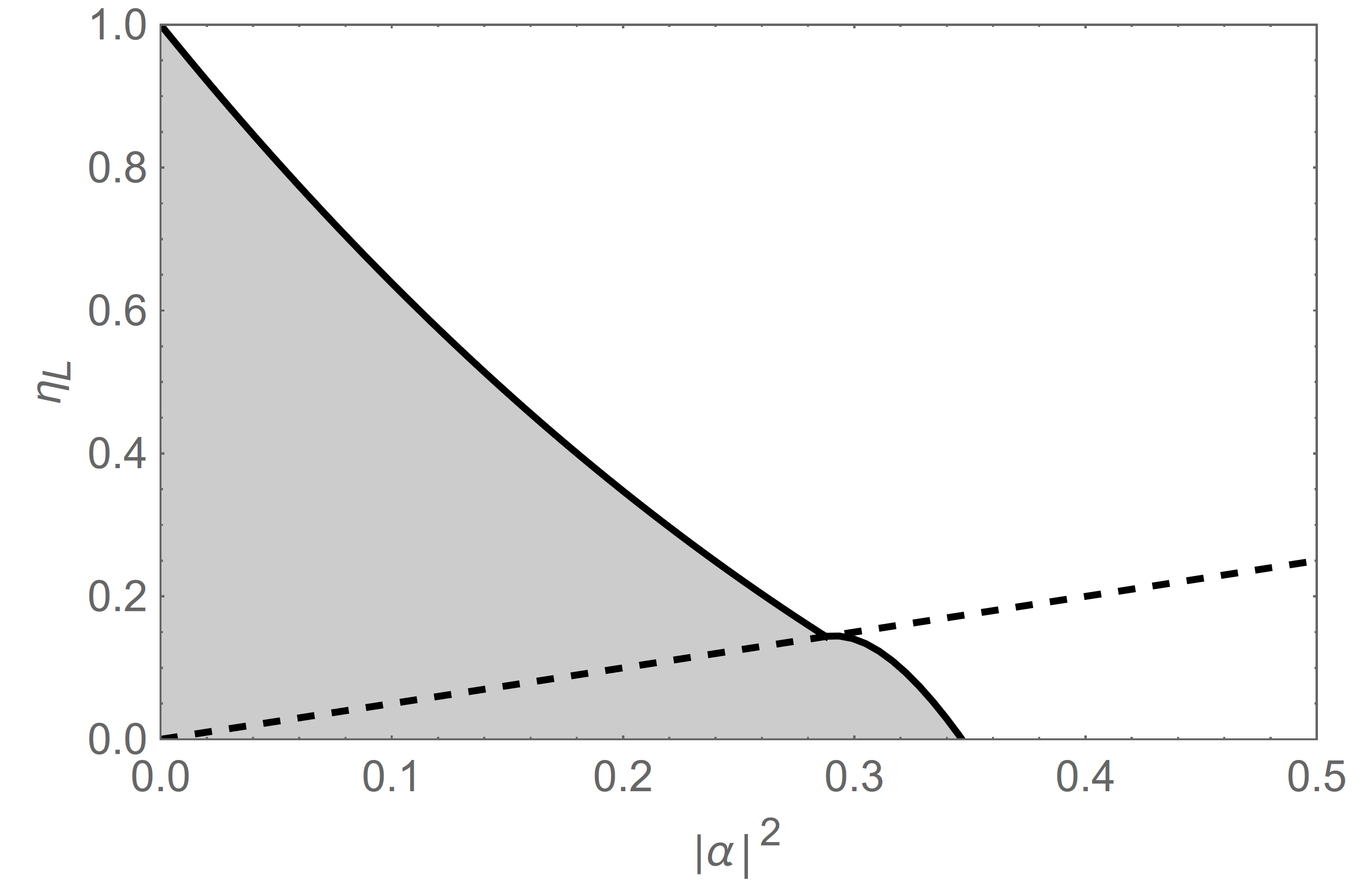}
        \caption{$\Delta\rightarrow0$}
        \label{fig:fig4f}
    \end{subfigure}

    \caption{Comparison of the information obtained by performing the proposed attack, denoted by $I$ and expressed in Eq.~\eqref{info} with the one obtained by performing a general collective attack, i.e., the Holevo bound, denoted by $\chi$ and expressed in Eq.~\eqref{holevo}, for several values of $\Delta$. The gray region denotes $I>\chi$, solid black line is described by $I=\chi$, black dashed line is $\eta_L=\frac{|\alpha|^2}{2}$}
\label{fig-holevocompdelta} 
\end{figure*}

\subsection{Subcarrier wave}
Subcarrier wave (SCW) QKD protocols \cite{miroshnichenko2018security} utilize phase-modulated states, prepared by a modulation with a harmonic electric signal $m\sin(\Omega t+\phi_j)$ applied to a coherent state $|\alpha_0\rangle_{\omega}$ at frequency $\omega$, of the following form:
\begin{gather}
    \bigotimes_{p=-S}^{S}|\alpha_0d^{S}_{0p}(\beta)e^{i\phi_jp}\rangle_{\omega+\Omega p},
    \end{gather}
where $d^{S}_{ij}(\beta)$ is the Wigner (small) d-matrix, $\beta$ is proportional to the modulation index $m$, $2S+1$ is the number of interacting frequency modes. Usually, $S\gg1$, and the Wigner d-matrix can be replaced with the Bessel function of the first kind:
\begin{gather}
    \lim_{S\rightarrow\infty} d^{S}_{ij}(\beta)=J_{i-j}(m).
\end{gather}
All sidebands contribute to the signal; its optical power we may define as follows:
\begin{gather}
    |\alpha|^2=|\alpha_0|^2(1-J^2_{0}(m)),
\end{gather}
and $|\alpha_0|^2J^2_{0}(m)$ can be considered as the optical power of the central frequency mode, which may be viewed as the reference beam. Usually, $m<1$, and the power of the reference is higher than the total power of the sidebands:
\begin{gather}
    J^2_{0}(m)>1-J^2_{0}(m).
\end{gather}
Then an eavesdropper may utilize the central frequency mode in order to augment each sideband with the part of the reference (by modulating it with some constant phase $\theta_e$) as follows:
\begin{gather}
    |\psi_{SCW}(\alpha,\theta_e,\phi_j)\rangle=\label{scwstate}\\
    =\bigotimes_{p\neq0}|\alpha_0J_{p}(m)e^{i\phi_jp}\rangle_{\omega+\Omega p}\otimes|\alpha_0J_{p}(m)e^{i\theta_e p}\rangle_{\omega+\Omega p}.\notag
\end{gather}
Then, for each pair of modes at frequency $\omega+\Omega p$, the projection~\eqref{project} is applied. Every estimation beyond this point is basically the same as in the general description of the attack.

\section{Results and discussion}\label{sec-result}

The article provides a thorough study of the proposed post-selective attack based on the PNS-inspired technique—multi-mode projection on the Fock subspace. The information known by an eavesdropper performing the attack described in the article can be estimated as in Eq.~\eqref{info}, which is defined only by three protocol characteristics: 
\begin{enumerate}
    \item Optical power (mean photon number) $|\alpha|^2$ of utilized coherent states;
    \item Informational basis structure—phase (half) difference  $\Delta=\frac{\phi_{(0)}-\phi_{(1)}}{2}$, where
    phases $\phi_{(0)}$ and $\phi_{(1)}$ correspond to bit values of $0$ and $1$ in a basis;
    \item Expected (without attack) optical attenuation in quantum channel $\eta_L$.
\end{enumerate}

Requirements for the attack to be performed are as follows:
\begin{enumerate}
    \item Utilized in a QKD protocol, quantum states can be transformed to the state of the form as in Eq.~\eqref{addingmode}, for instance, as in Eqs.~\eqref{mzistate}, \eqref{ptstate1}, \eqref{ptstate2}, \eqref{pmstate} or \eqref{scwstate};
    \item A protocol should not implement decoy states and/or single-photon fraction extraction, for instance, as in GLLP framework \cite{gottesman2004security}.
\end{enumerate}
The efficiency of the attack has been studied by comparison with the Holevo bound, as in \cite{kronberg2025generalization}. For a wide range of protocol parameters, specifically $|\alpha|^2$ and $\eta_L$, and commonly used value of $\Delta=\pi/2$, high efficiency of the attack is confirmed; the region of the attack's superiority compared to collective attacks is shown in Fig.~\ref{fig:holevocomp}. At this point we conclude that the attack poses a significant threat and should be addressed with proper countermeasures.

We admit that countermeasures should be considered individually for a particular protocol in order to take into consideration various features. However, the analysis of the attack allows picturing some general ideas, which can be considered as a foundation for future examinations in regard to countermeasure development; these are as follows:
\begin{enumerate}
    \item Implementation of decoy states \cite{lo2005decoy,ma2005practical,wang2005beating} or similar ideas, i.e., utilization of several values of $|\alpha|^2$, expands maintain-detection-rate condition~\eqref{cond1}, that, in turn, can make it impossible (or at least significantly limits) for an eavesdropper to maintain several expected detection rates at once;
    \item A similar idea as an utilization of several values of mean photon number, $|\alpha|^2$, can be a modification of the detection scheme in a way, that legitimate users monitor not only detection rate proportional to a mean photon number, but also some correlation function proportional to the square of a mean photon number; it expands maintain-detection-rate condition~\eqref{cond1} as well and makes it impossible (or at least significantly limits) for an eavesdropper to maintain both at once;
    \item Single-photon fraction extraction \cite{gottesman2004security} during post-processing by legitimate users may break correlations between an eavesdropper and a key due to a lack of multi-photon events that may cause these correlations;
    \item The last but not least possible solution is to directly examine Eq.~\eqref{info} and tweak parameters that reduce the area of the attack superiority compared to collective attacks (the Holevo bound); we elaborate on that in more detail further.
\end{enumerate}

One of the options to decrease the amount of information an eavesdropper may obtain by performing the attack is to modify the protocol in a way that $\Delta$ can be decreased; results can be seen in Fig.~\ref{fig-holevocompdelta}.  Observe that even if $\Delta\rightarrow0$, there is still a region where $I>\chi$. Consider $\lim_{\Delta\rightarrow0}\frac{I}{\chi}\ge1$ at $\eta_L=0$, where we use the following relation
\begin{gather}
    \lim_{\Delta\rightarrow 0}\frac{\chi^{(n)}(\Delta)}{\chi}=\frac{n}{4|\alpha|^2},
\end{gather}

then $|\alpha|^2\le\ln(2)/2$. The exact limiting value of $|\alpha|^2$ is close to the latter; however, it can be found only numerically; its approximate value is $|\alpha|^2\approx0.34044$ at $\Delta\approx0.0237\cdot\pi$. At this point, regardless of the value of $\Delta$, in the region $|\alpha|^2\gtrapprox 0.34044$, the proposed attack is less efficient compared to collective attacks, and legitimate users still may use the Holevo bound as the upper limit for estimations of the intercepted information.

\section*{Acknowledgements}
Kozubov A. and Gaidash A. acknowledge support from the Russian Science
Foundation (Project No. 25-71-10034).

\appendix
\section{Holevo bound}\label{app1}
For any non-orthogonal states $|\psi(\alpha,\phi_{(0)})\rangle$ and $|\psi(\alpha,\phi_{(1)})\rangle$, an orthonormal basis $|u_0\rangle$ and $|u_1\rangle$ can be obtained by the Gram-Schmidt process:
\begin{gather}
    |u_0\rangle=|\psi(\alpha,\phi_{(0)})\rangle,\\
    |u_1\rangle=\frac{|\psi(\alpha,\phi_{(1)})\rangle-c|\psi(\alpha,\phi_{(0)})\rangle}{\sqrt{1-|c|^2}},
\end{gather}
where $c=\langle\psi(\alpha,\phi_{(1)})|\psi(\alpha,\phi_{(0)})\rangle$, and, in turn, these states can be defined in terms of the found basis:
\begin{gather}
    |\psi(\alpha,\phi_{(0)})\rangle=|u_0\rangle,\\
    |\psi(\alpha,\phi_{(1)})\rangle=c|u_0\rangle+\sqrt{1-|c|^2}|u_1\rangle.
\end{gather}
An ensemble density matrix is then given by
\begin{gather}
    \rho=\frac{1}{2}\Big(|\psi_0\rangle\langle\psi_0|+|\psi_1\rangle\langle\psi_1|\Big)\notag\\
    =\frac{1}{2}\begin{pmatrix}
        1+|c|^2 & c^*\sqrt{1-|c|^2}\\
        c\sqrt{1-|c|^2} & 1-|c|^2
    \end{pmatrix},
\end{gather}
and its eigenvalues are as follows:
\begin{gather}
    \lambda_{\pm}=\frac{1\pm|c|}{2}.
\end{gather}
The Holevo bound can be expressed as the von Neumann entropy of a given ensemble density matrix or, equivalently, the Shannon entropy of its eigenvalues:
\begin{gather}
    \chi=-\sum_{i}\lambda_i\log_2(\lambda_i)=h\Big(\frac{1-|c|}{2}\Big),
\end{gather}
and
\begin{gather}
    |c|=e^{-|\alpha|^2(1-\cos(2\Delta))}.
\end{gather}
\bibliography{apssamp}

\end{document}